\documentclass[epj,referee]{svjour}

\usepackage{graphicx}
\usepackage{amsmath}

\newcommand{\beq}{\begin{equation}}
\newcommand{\enq}{\end{equation}}
\newcommand{\bea}{\begin{eqnarray}}
\newcommand{\ena}{\end{eqnarray}}
\newcommand{\ad}{a^{\dagger}}
\newcommand{\rr}{{\mathbf r}}
\newcommand{\rhos}{\rho_{\rm s}}
\newcommand{\rhoxy}{\rho_{\rm s,XY}}
\newcommand{\kl}{k_{\rm L}}
\newcommand{\tc}{t_{y\rm c}}
\newcommand{\ttc}{\tilde{t}_{y\rm c}}
\newcommand{\dd}{\mathrm{d}}

\newcommand{\thecitation}{\cite{ho2004,cazalilla2006,gangardt2006} }

\newcommand{\refeq}[1]{Eq.~(\ref{#1})}

\newcommand{\showfig}[1]{\includegraphics[width=0.44\textwidth]{#1}}

\begin{document}
\title{One-dimensional phase transitions in a two-dimensional optical lattice}
\author{Magnus Rehn\inst{1,2} \and 
Sara Bergkvist\inst{3} \and 
Anders Rosengren\inst{3,4} \and
Robert Saers\inst{1} \and
Martin Zel{\'a}n\inst{1} \and
Emil Lundh\inst{1} \and
Anders Kastberg\inst{1}
}
\institute{Department of Physics, Ume{\aa} University, 
SE-90187 Ume{\aa}, Sweden 
\and
Quantum Research Group, School of Physics, Westville
University of KwaZulu-Natal, Durban, 4041, South Africa 
\and
Condensed Matter Theory, Department of Theoretical Physics, AlbaNova University Center, KTH, SE-106 91 Stockholm, Sweden 
\and
NORDITA, AlbaNova University Center, SE-106 91 Stockholm, Sweden}

%\date{\today}

\abstract{
A phase transition for bosonic atoms in a two-dimensional anisotropic 
optical lattice is considered. 
If the tunnelling rates in two directions are different, 
the system can undergo a transition between 
a two-dimensional superfluid
and 
a one-dimensional Mott insulating array of strongly coupled tubes. 
The connection to other lattice models is exploited in 
order to better understand the phase transition. 
Critical properties are obtained using quantum Monte Carlo calculations. 
These critical properties are related to correlation properties of the bosons 
and a criterion for commensurate filling is established.
}

\maketitle

\section{Introduction}
Atoms in optical lattices offer new versatile ways of studying many-body 
phenomena \cite{jessen1996,bloch2005}. 
Interference patterns of laser light provide periodic 
potentials for atoms, which are cooled to the nanokelvin range 
using laser and evaporative cooling. 
In this way, ground-state properties of these interesting many-body 
systems may be probed using a variety of optical techniques. 

The optical potentials can be controlled with great range and precision, 
giving access to any kind of Bravais lattice as well as superlattices 
and quasiperiodic potentials \cite{grynberg2001}. 
The geometrical setup and choice of polarisation determine 
the lattice geometry, and the laser irradiance determines the amplitude 
of the potential. If the laser light is detuned to the red of the atomic 
transition, the potential will typically form an array of wells in which 
the atoms may be trapped if the potential is deep enough.
The height and width of the potential barriers separating the wells determine the 
rate of tunnelling between them. In this way, the dimensionality of the 
sample can be controlled: 
%EL
making the potential barriers high enough in 
one Cartesian direction will effectively inhibit all tunnelling and 
result in a stack of independent, two-dimensional (2D) lattices. 
Increasing the potential along a second direction yields a 2D array of 
1D tubes of atoms.

Recently, such dimensional crossovers in optical lattices and their 
relation to other theoretical models have attracted 
considerable attention. 
References \thecitation
studied a 2D 
array of one-dimensional tubes of atoms, known as 
Tomonaga-Luttinger liquids (TLL) \cite{giamarchi2004}. 
In a random-phase approximation, 
phase boundaries, coherence properties and excitation spectra were 
expressed in terms of the parameters that characterise the TLLs. 
In Ref.\ \cite{bergkvist2007}, we studied the corresponding situation 
in a 2D lattice geometry and investigated the relation between this 
model and a classical XY model 
%EL
\cite{sachdev1999}. It was determined that 
the transition 
between independent TLL tubes and a 2D superfluid is of 
Berezinskii-Kosterlitz-Thouless (BKT) type 
\cite{melko2004,weber1988,olsson1995b} 
if the density is held fixed, and as an example, 
the critical point was determined for a specific value of tunnelling matrix 
element and filling. 

In this paper, we further investigate the 
%EL
transitions in the 2D case. 
Using quantum Monte Carlo calculations, we investigate 
the link between coherence properties and the critical point for the 
phase transition. We confirm 
%EL
%expectations 
that the transition is 
determined by a TLL theory and that it depends crucially on the 
commensurability of the atoms in the lattice. In Sec.\ \ref{sec:theory}, 
we lay out the theory. In Sec.\ \ref{sec:mapping}, we show how the 
present Hamiltonian is equivalent to other lattice models, which 
clarifies the nature of the phase transition. 
Section \ref{sec:scaling} explains the numerical method and the 
finite-size scaling performed in 
order to locate the phase transition, and presents the results 
for the phase transition at unit filling.
In Sec.\ \ref{sec:correlations} we show how it is linked 
to the coherence properties. In Sec.\ \ref{sec:doubletransition}, 
it is illustrated how changing one of the parameters of the theory
allows for a crossing a two subsequent phase transitions.
In Sec.\ \ref{sec:densities}, we 
investigate commensurability and density dependence, 
and finally in
Sec.\ \ref{sec:conclusion} we summarise and conclude.

\section{Phase transitions in a 2D Hubbard model}
\label{sec:theory}
Consider a one-component gas of bosonic atoms in a 3D lattice potential created by 
three standing waves at right angles,
\beq
V(\rr) = V_{0x}\cos^2\kl x + V_{0y}\cos^2\kl y+V_{0z}\cos^2\kl z,
\enq 
where $\kl=2\pi/\lambda$ is the angular wave number of the light, and 
the potential heights $V_{0x}, V_{0y}$ and $V_{0z}$ are determined by 
the laser irradiance and detuning. 
In the tight-binding approximation, the many-body Hamiltonian is 
expanded in a basis of Wannier functions $w_n(\rr-\rr_i)$, 
each of which is localised in a well centred at position $\rr_i$, 
where $i$ labels the wells and $n$ is a band index. 
If the 
%EL
wells are 
tight enough, and the 
atoms are cold enough, only the lowest-energy Wannier function within 
each well, $w_0(\rr-\rr_i)$, needs to be taken into account; this is 
the lowest Bloch band.
The gas is then described by the bosonic single-band Hubbard model
\cite{jaksch1998},
\beq
H = -\sum_{<ij>} t_{ij} (a_i \ad_{j}+{\rm h.c.}) + \frac{U}{2}\sum_i \ad_i\ad_i a_i a_i 
- \mu\sum_i \ad_i a_i,
\enq
where $i$ is a placeholder for the three indices $i_x,i_y,i_z$ enumerating 
the lattice sites in the Cartesian directions, and the sum subscripted 
$<\!\!ij\!\!>$ runs over pairs of neighbouring sites. The parameter $U$ is 
the on-site interaction strength, $t_{ij}$ is the tunnelling matrix 
element for the barriers between sites $i$ and $j$,
and $\mu$ is the chemical potential, adjusted in the calculations 
to  give the desired density of atoms. 
The on-site interaction strength is defined as 
\beq
U = \frac{2\pi\hbar^2 a}{m}\int \dd^3r |w_0(\rr-\rr_i)|^4,
\enq
where $a$ is the s-wave scattering length of the atoms and $m$ is the 
atomic  mass. 
Since this optical lattice is created by three 
independent 
pairs of standing waves, 
the tunnelling matrix element $t_{ij}$ in one particular term 
takes on one of three values. 
If the wells $i$ and $j$ are neighbours along the $x$ direction, its value is 
\beq
t_x = \int \dd^3\rr w_0^*(x,y,z)\left[-\frac{\hbar^2}{2m}\nabla^2 + V(\rr)
\right]w_0(x+d,y,z),
\enq
where $d=\lambda/2$ is the lattice spacing. The tunnelling matrix 
elements $t_y$ and $t_z$ are defined similarly.

\begin{figure}
\showfig{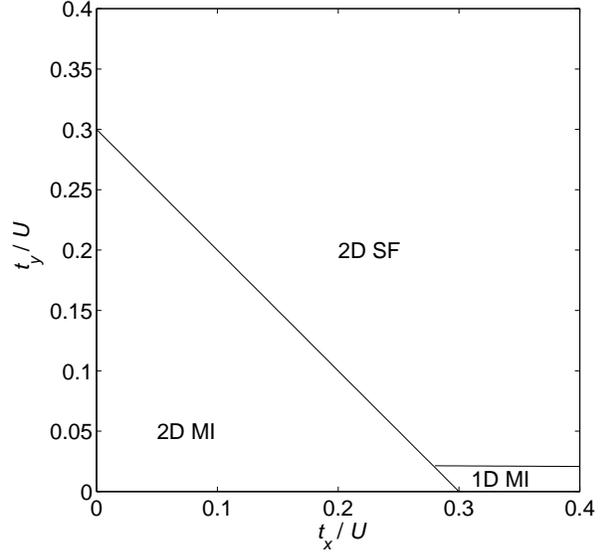}
\caption{Predicted phases of the anisotropic Hubbard model in two dimensions. 
``2D MI'' and ``2D SF'' stand for the two-dimensional Mott insulating phase 
and superfluid phase, respectively. The phase ``1D MI'' is predicted to 
exist only when one of the side lengths of the system is 
finite. In this figure, the system extends infinitely along the $y$ direction 
and is 4 sites wide in the $x$ direction. 
The slight decrease of the corresponding phase boundary is not visible in 
the figure.
}
\label{fig:phasediagram}
\end{figure}

It turns out that the tunnelling matrix elements depend exponentially on 
the potential depth \cite{jaksch1998}. 
As a result, one can easily make $t_x$, $t_y$ 
and $t_z$ differ by orders of magnitude by a judicious choice of the laser 
irradiances. Thus, $V_{0z}$ can be chosen large 
enough that $t_z$ can be neglected and the sample is effectively 2D. 
In the following, we confine the discussion to the 2D Bose-Hubbard model 
(henceforth referred to as the Hubbard model for brevity). 
The phase diagram is sketched in 
Fig.~\ref{fig:phasediagram} and we now describe the general features.
When both $t_x$ and $t_y$ are small compared with $U$, 
and in addition the number of bosons is commensurate with the number of lattice sites, 
the sample 
is Mott insulating (MI), with exponentially decaying phase correlations and 
suppressed fluctuations in the number of particles per site. 
In the opposite limit, when $t_x$ and $t_y$ are of the same order as $U$ 
or larger, the 
ground state is the 2D superfluid (2D SF) state, characterised by 
long-range phase coherence in 2D and a fluctuating number of particles 
per site.
In one dimension with one particle per site, there is a
quantum phase transition between these phases at $t/U=0.3$ 
\cite{kuhner2000}. 
%EL
In higher dimensions, one may apply a mean-field approximation
\cite{sachdev1999}, dictating 
that 
the transition occurs when the sum of all 
matrix elements equals the 1D value, {\it i.e.},
\beq
\frac{t_x+t_y}{U} = 0.3
\enq
in two dimensions.
The exact, numerically obtained 2D result \cite{freericks1996} 
deviates slightly from this 
mean-field result. 
Furthermore, quantum Monte Carlo 
calculations 
performed on finite systems tend to underestimate the critical value 
for the phase transition.

If one of the tunnelling matrix elements, say $t_x$, is relatively 
large and the other, $t_y$, is small, there may in fact under certain 
circumstances exist 
a phase where 
there is superflow along one dimension but not along the other. Such 
a state will be called the 1D MI phase. It turns out that in an 
{\em infinite} 2D 
system, such a state is absent: if there is superflow in the $x$ 
direction, then any nonzero tunnelling along the $y$ 
direction will put the sample into the 2D SF state \cite{efetov1974} 
(as a corollary, any finite tunnelling along the $z$ direction will result 
in a 3D SF state). However, as was first noted by Ho {\it et al.} 
and Gangardt {\it et al.}
\thecitation (although applied to the 3D case), 
the situation changes when the 
sample is {\em finite} along the strongly coupled $x$ direction. 
The system can then be thought of as an array 
of finite tubes lining up along the $y$ direction and  
extending along the $x$ direction. 

Because 
of the finite excitation energy within the tubes, there can now exist a 
1D MI phase if $t_y$ is sufficiently small. By the same argument, 
the tunnelling along the $z$ direction can be neglected, as we 
assumed.

In a finite system, there is no true phase transition. 
In practice, this is not a problem since the phase transition is replaced 
by a crossover. In theory, it is always possible to let the system 
extend indefinitely along the weakly coupled $y$ direction, while 
keeping it finite along 
the $x$ direction. There is then a true phase transition between 
insulating and superfluid behaviour along the $y$ direction. 
We call this the 1D~MI - 2D~SF transition.

In Refs.\ \thecitation, the 1D tubes were 
described using TLL theory. 
This theory describes a 1D many-body system, which is characterised 
by two parameters independently of statistics and 
the detailed properties of the constituent particles, namely the 
sound velocity $v_{\rm s}$ and the TLL parameter $K$, defined as 
\beq
K=\frac{v_{\rm F}}{v_{\rm s}}.
\enq 
Here, $v_{\rm F}=\hbar\pi\rho/(md)$, where $\rho$ is the filling factor, 
{\it i.e.}, the number of bosons per site, and $d$ is the lattice constant.
The TLL parameter $K$ determines the behaviour of the 
particle-particle correlations $\Gamma(i,j)$, 
which obey the power law
\beq
\label{correlation}
\Gamma(i,j) \equiv \langle \ad_i a_j\rangle \propto |\rr_i-\rr_j|^{-1/(2K)}.
\enq
For the discussion of the correlations along different 
directions, we introduce the notation $\Gamma_x$ and $K_x$ for the 
correlations within the strongly coupled tubes, as well as $\Gamma_y$ 
and $K_y$ for describing correlations along the weakly coupled $y$ 
direction. 

Since a 
%EL
larger 
$K_x$ implies greater coherence along the $x$ direction, 
we expect $K_x$ to increase with $t_x/U$. Furthermore, it is a known 
exact result that 
\beq
2 \leq K_x < \infty
\enq
for a system of lattice bosons with only on-site interactions 
\cite{giamarchi2004}. If one further decreases the tunnelling 
beyond 
the point where $K_x=2$, one enters the MI phase. However, the 
quantitative relation between $t_x$ and $K_x$ is not known in closed 
form.

\section{Equivalence to other lattice models}
\label{sec:mapping}
In order to obtain an expression for the transition point, 
the TLL tubes were in 
Refs.\ \cite{ho2004,cazalilla2006} treated as structureless 
sites by integrating out all degrees of freedom except a number 
%EL
operator $N_j$ and a 
phase operator 
$\phi_j$ for each tube 
(the analysis was done with a 3D system in mind, but the results 
apply to the 2D case as well).
In that way, a 
number-phase model \cite{vanotterlo1995}
was obtained with a governing Hamiltonian
\beq
H = -E_{\rm J}\sum_{<ij>}\cos(\phi_i-\phi_{j}) +E_{\rm C}\sum_i(N_i-N_0)^2,
\enq
where $N_0$ is the equilibrium number of particles per tube, and 
$<\!\!ij\!\!>$ denotes neighbouring tubes in the $y$ direction. $E_{\rm J}$ 
is usually referred to as the tunnelling energy and $E_{\rm C}$ as the 
charging energy.
The mapping is expected to hold in the 1D~MI phase, but above the phase 
transition, where the whole 2D system is superfluid, it may not be 
valid.
It was found in Refs.\ \cite{ho2004,cazalilla2006} that 
\beq
E_{\rm J}=t_y N_0^{1-\alpha_x},
\enq
and
\beq
\label{ec}
E_{\rm C} = \frac{C_0U}{L_x},
\enq
where the exponent
\beq
\alpha_x = \frac{1}{2K_x}.
\enq
Thus, it is predicted that the behaviour of the particle-particle 
correlation function within the tubes determines the location of the 
critical point for the decoupling of the tubes.
Here, $L_x$ is the number of sites that the lattice extends 
in the $x$ direction, and $C_0$ is a constant 
that can in general not be obtained in closed form.
The number-phase Hamiltonian undergoes a SF-MI quantum phase transition 
at the critical point where
\beq
\frac{E_{\rm J}}{E_{\rm C}} = {\rm constant},
\enq
where the right-hand side is a universal numerical constant to be determined. 
Hence, for the critical value 
$\tc$
of $t_y$,
\beq
\label{crit_prop}
\frac{\tc}{U} \propto \rho^{-1+\alpha_x} L_x^{-2+\alpha_x},
\enq
as long as the analysis of Ref.\ \cite{ho2004,cazalilla2006} provides a 
valid model for the anisotropic optical lattice. The same functional form 
for the critical point was obtained in Ref.\ \cite{gangardt2006} by means 
of a random-phase approximation. 
In Eq.~(\ref{crit_prop}), the filling $\rho$ is defined as the 
mean number of atoms per lattice site, and is thus dimensionless.
Due to the known constraints on $K_x$, the power on $L_x$ lies between 
-1.75 and -2, where the former holds close to the transition to a 2D MI.
We define the effective coupling constant
\beq
\tilde{t}_y = \frac{t_y}{U} L_x^{2-\alpha_x}\rho^{1-\alpha_x}.
\enq
Stated in terms of this quantity, the critical value of the tunnelling 
$t_y$ can be written as
\beq
\label{transnphi}
\frac{\tc}{U} = \ttc L_x^{-2+\alpha_x}\rho^{-1+\alpha_x},
\enq
where $\alpha_x$ and $\ttc$ are constants to be determined.

The nature of the quantum phase transition in the number-phase model 
can be understood by 
making use of the general result that a quantum phase transition in $D$ 
dimensions is in the same universality class -- {\it i.e.}, it has the same 
critical properties -- as a classical phase transition in a certain 
corresponding model 
in $D+1$ dimensions \cite{sachdev1999}. 
In 1D, with the 
number of bosons kept fixed, the transition studied here is in the same 
universality class as the classical XY model in 2 dimensions at 
a finite temperature $T_{\rm XY}=1/(k_{\rm B}\beta_{\rm XY})$. 
The mapping is accomplished by identifying \cite{sondhi1997}
\bea
\beta_{\rm XY} &=& \sqrt{\frac{E_{\rm J}}{E_{\rm C}}},\nonumber\\
L_{x,\rm XY} &=& \beta\sqrt{E_{\rm J}E_{\rm C}},\nonumber\\
L_{y,\rm XY} &=& L_y,
\ena
where quantities with the subscript XY refer to the 2D XY model and 
quantities without
that subscript refer to the number-phase model. 
%The XY-model temperature is made dimensionless by defining the XY 
%coupling strength to be the unit of energy, as is customary \cite{sondhi1997}.
In terms of the underlying anisotropic Hubbard model, we can 
thus write 
\bea
\beta_{\rm XY} &= & \left(\frac{t_y L_x^{2-\alpha_x}\rho^{1-\alpha_x}}{C_0U}\right)^{1/2},\nonumber\\
L_{x,\rm XY} &= & \beta\left(C_0 U t_y L_x^{-\alpha_x}\rho^{1-\alpha_x}\right)^{1/2},\nonumber\\
L_{y,\rm XY} &=& L_y,
\label{mapping2}
\ena
where $C_0$ is the same constant as in Eq.~(\ref{ec}), 
and $\beta=1/(k_{\rm B}T)$, where $T$ is the temperature of the Hubbard model. 
This mapping shows that if the XY model has a phase transition at 
an inverse temperature $\beta_{\rm XY}$, then the Hubbard model has a 
quantum phase transition when $\tilde{t}_y$ reaches a critical value, 
consistent with Eq.~(\ref{transnphi}). The critical point will 
depend on the tunnelling within the tubes, $t_x$, through the 
Luttinger parameter $K_x$ and also through the unknown constant 
of proportionality $C_0$ in Eq.~(\ref{mapping2}).

%EL
%In the present case, the 1D number-phase model derived from the 2D 
%anisotropic Hubbard model can be mapped onto the 2D XY model if the density 
%is kept fixed. This model 
The 2D XY model
exhibits a BKT transition 
\cite{melko2004,fisher1989,krauth1991,kuhner1998}. 
This kind of transition 
possesses several characteristics, among them a universal 
jump in the superfluid density at the transition point. 
According to the series 
of mappings performed here, from a 2D anisotropic Hubbard model via 
a 1D number-phase model to a 2D XY model, the anisotropic 
Hubbard model exhibits a BKT transition. The 
findings of Ref.\ \cite{bergkvist2007} 
confirm these
expectations, and the present paper further expands on the subject.

\section{Finite size scaling}
\label{sec:scaling}
In the numerical calculations, the 2D Hubbard model was 
%EL
%solved 
simulated
using the 
stochastic series expansion method \cite{syljuasen2002,syljuasen2003}. 
The %on-site interaction strength $U$ was set to 1, and the 
chemical potential $\mu$ was tuned to obtain the desired 
mean number of particles. By selecting those 
%EL
Monte Carlo 
steps that correspond 
to a 
%EL
%certain 
fixed
number of particles, it was made sure that the calculations 
were performed at a given density, which is important 
for the characteristics of the phase transition. We chose 
$\beta=1000U^{-1}$ in order to make sure that 
ground-state properties were calculated, and the number of states per 
site was chosen to 6 in order to ensure convergence.

\begin{figure}
\showfig{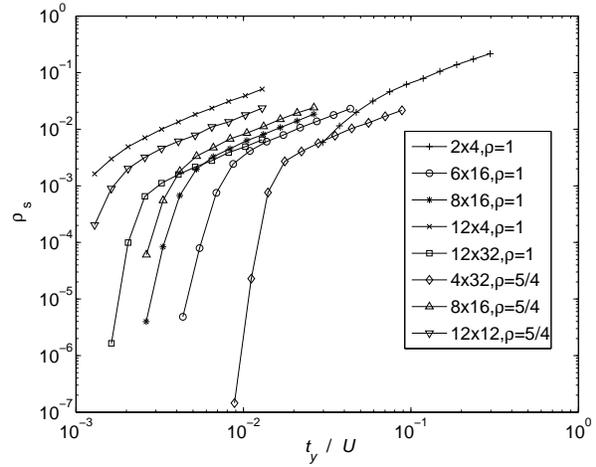}
\caption{A few arbitrarily selected curves showing the superfluid 
density as a function of 
the ratio of tunnelling $t_y$ to on-site interaction $U$. The different 
curves correspond to different data sets obtained at different fixed 
side lengths $L_x$ and $L_y$, and different filling factors $\rho$.}
\label{fig:rhos_unscaled}
\end{figure}

Systems of $L_x\times L_y$ sites were simulated, where the side lengths 
$L_x$ and $L_y$ were varied in order to assess the predicted dependence 
of the critical point on the length discussed in Sec.~\ref{sec:theory}. 
The boundary conditions were chosen to be periodic, which is necessary 
for calculating the superfluid density as will be described. 
The most important calculated quantities are the 
superfluid density $\rhos$ and the particle-particle correlations 
$\Gamma_x$ and $\Gamma_y$. The superfluid density is numerically 
computed via the winding number $W_y$ as 
\beq
\rhos = \frac{\langle W_y^2\rangle L_y}{L_x \beta},
\enq 
where $W_y$ is the net number of times that a particle 
line crosses the periodic boundary in the $y$ direction in the 
simulations \cite{melko2004}. 
The superfluid density as a function of the ratio of tunnelling matrix element 
and on-site interaction energy, $t_y/U$, is plotted in 
Fig.~\ref{fig:rhos_unscaled} for a few examples of parameter values. 
It is clear that the curves make a sharp drop at a transition point 
when $t_y/U$ becomes small enough. 
The transition point seems to depend on both $L_x$ and $\rho$, 
as was predicted in Sec.~\ref{sec:mapping}, and we now turn to the 
problem of calculating this dependence.

In order to locate the critical point, 
finite-size scaling must be performed 
\cite{weber1988,olsson1995b,melko2004}. 
However, the analysis of the present problem presents several 
difficulties compared with a classical XY model. 
At the BKT transition, the superfluid density assumes 
a value known as the universal jump, proportional to the critical temperature. 
This is routinely used in 2D~XY-model simulations, but since 
in the present case the two quantities are only known to within an unknown 
constant, we cannot make use of this relation. 
Moreover, the mapping between the Hubbard model and the 2D~XY model is only 
valid at and below the phase transition. In the 2D~SF phase, the 
coherence between tubes may be comparable to that within the tubes and 
the mapping is not valid. 
Finally, a quantum Monte Carlo calculation of the 2D Hubbard model is 
very time consuming and in the parameter regime of interest, where the 
parameters $t_x$ and $t_y$ differ by orders of magnitude, it is 
hard to obtain data with high accuracy. This is aggravated by the fact 
that the side length $L_y$ varies between 4 and 32 in the simulations. 

As  seen in Sec.\ \ref{sec:mapping}, the anisotropic Hubbard 
model is predicted to have the same properties close to the transition 
as the 2D XY model has close to the BKT transition, if the 2D XY model 
has a superfluid density given by
\beq
\rhoxy = \frac{\langle W_y^2\rangle L_y}{\beta t_y 
\rho^{1-\alpha_x} L_x^{1-\alpha_x}}.
\enq
The BKT transition occurs when $\tilde{t}_y$ 
%=t_yL_x^{2-\alpha_x}\rho^{1-\alpha_x}/U$ 
assumes a critical value; at this point, all the curves 
$\rhoxy(\tilde{t}_y;L_x;L_y)$ 
computed for different parameter values 
should ideally coincide (taking finite-size effects into account). 
This means that $\alpha_x$ has to be optimised so that all the curves for 
different parameter values coincide as closely as possible. 
This is accomplished by first considering sets of data series 
with a given $L_y$ and different $L_x$. 
If the analysis in Sec.\ \ref{sec:mapping} is correct, then all systems 
of size $L_x\times L_y$ can be approximated as 1D chains of length $L_y$, 
and therefore the results for $\rhoxy$ as a function of $\tilde{t}_y$ 
should coincide 
between data series with similar $L_y$ and different $L_x$, if only 
the parameter $\alpha_x$ is chosen correctly. 
As noted in 
Sec.\ \ref{sec:mapping}, the correspondence is only expected to hold 
in the 1D MI phase, so that for $\tilde{t}_y>\ttc$, one cannot 
require the data to coincide. 
Thus, one needs to consider the variance among the curves below a 
supposed critical point, 
and choose the value of $\alpha_x$ that minimises the variance. This is to be 
done for each $L_y$ separately, and then the results for different $L_y$ 
can be compared. 
We thus do a spline interpolation of the points over the 
relevant range of $\tilde{t}_y$ and compute, for a given $L_y$ and $\alpha_x$,
\begin{align}
&{\rm Var}(\rhoxy)(\tilde{t}_y,\alpha_x,L_y) =  \nonumber\\
&\sum_{L_x} \left(\rhoxy(\tilde{t}_y;L_x;L_y) - 
\frac1{N_{L_x}}\sum_{L_x}\rhoxy(\tilde{t}_y;L_x;L_y)\right)^2, 
\label{variance}
\end{align}
and furthermore the summed variance over the whole range of $\tilde{t}_y$ is
\beq
\label{variance2}
\overline{\rm Var}(\rhoxy)(\alpha_x,L_y) = 
\frac{\int \dd\tilde{t}_y {\rm Var}(\rhoxy)(\tilde{t}_y,\alpha_x,L_y)}{\int \dd\tilde{t}_y1}. 
\enq
In Eqs.~(\ref{variance}-\ref{variance2}), 
$N_{L_x}$ is defined as the number of different values of $L_x$ used.
Since the curves are supposed to coincide only below the critical point, 
the integration limits for $\tilde{t}_y$ are chosen as 
$0 \leq \tilde{t}_y \leq 0.3$, anticipating the result that the critical 
point is close to $\ttc=0.3$. The result does not depend 
strongly on the chosen integration limits. 
\begin{figure}
\showfig{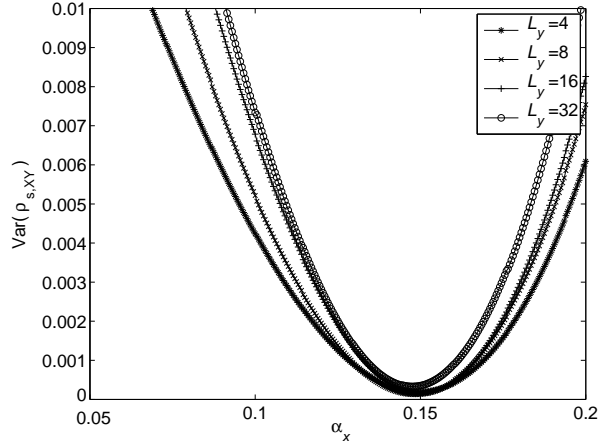}
\caption{Variance in the superfluid density among data sets 
as a function of the unknown exponent $\alpha_x=1/(2K_x)$. 
}
\label{fig:skalfig_d1}
\end{figure}
For each value of $L_y$, we obtain an optimal value for $\alpha_x$. 
Averaging over different $L_y$, the result is
\beq
\label{alpha_scal}
\alpha_x = 0.15,
\enq
resulting in
\beq
K_x = 3.4.
\enq

When data for {\em different} system sizes $L_y$ are compared, they are 
expected to coincide at the critical point, but not below or above. 
In addition, the coincidence of the curves is exact only in the limit 
$L_y\to\infty$, but it is known how to make the 
lowest-order correction for finite $L_y$. 
At the BKT transition point in the 2D XY model, the superfluid density 
depends asymptotically on the size of the finite sample as 
\beq
\label{weberminnhagen}
\rhoxy(\infty) = \frac{\rhoxy(L_y)}{
1 + \frac{1}{2\ln(L_y)+C}},
\enq
where $\rhoxy(L_y)$ is the superfluid density computed using a side 
length $L_y$.
This result is known as Weber-Minnhagen scaling \cite{weber1988}, 
and in Ref.\ \cite{melko2004} it was found that the procedure applies 
to non-quadratic systems as long as the winding number is 
computed along the shorter dimension of the sample. In our effective 
XY model, $L_{x,\rm XY}$ is proportional to the inverse temperature of 
the Hubbard model and it is thus much larger than $L_y$. 
The constant $C$ was in Ref.~\cite{olsson1995b} found to be equal to 
1.8.

\begin{figure}
\showfig{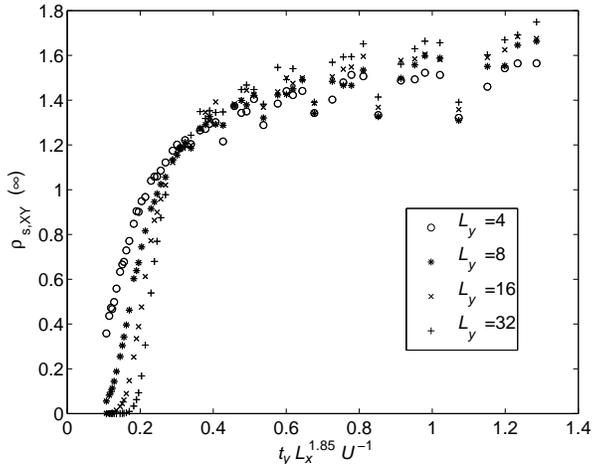}
\caption{Superfluid density for the dual XY model, scaled using the 
best fit parameter $K_x=3.4$, for a filling $\rho=1$. The plot contains all 
the data sets used in the scaling. The values of $L_y$ are as indicated in 
the legend.
}
\label{fig:collapsed}
\end{figure}

Figure \ref{fig:collapsed} collects the simulated data for all different 
values of $L_x$ and $L_y$. Ideally, all the data is expected to coincide
at the critical point $\ttc$. 
The variance among these curves is now recorded as a function of $\tilde{t}_y$ 
and the smallest variance is at 
\beq
\ttc=0.33.
\enq
Table \ref{tab:critical} summarises the computed parameters for a few 
different values of tunnelling $t_x$ and filling $\rho$. The constant 
$\ttc$ was defined in Eq.~(\ref{transnphi}), the exponent 
$\alpha_x$ and the 
result for $K_x$ are as obtained above, and the value for $K_x$ obtained 
from correlations will be discussed in Sec.~\ref{sec:correlations}.
\begin{table}
\caption[t1]{\label{tab:critical}
Parameters determining the 2D SF-1D MI transition. The method to 
calculate the error bars are described in the text. The data for 
$t_x/U=0.3$ are lifted from Ref.~\cite{bergkvist2007}.
}
\begin{tabular}{c|c|c|c|c|c}
\hline
\hline
$t_x/U$ & $\rho$ & $\ttc$ & $\alpha_x$ & $K_x$ & $K_x$ \\
& & & & (transition) & (correlation) \\
\hline
0.3 & 1.0 & 0.3 & 0.25 & 2.0 & -- \\
0.5 & 1.0 & 0.33  & 0.15 & 3.4 & 3.0  \\
0.5 & 5/4 & 0.32 & 0.15 & 3.4  & 2.7  \\
0.5 & 19/16 & - & -     & -    & 2.1 \\
\hline
\hline
\end{tabular}
\end{table}

\begin{figure}
\showfig{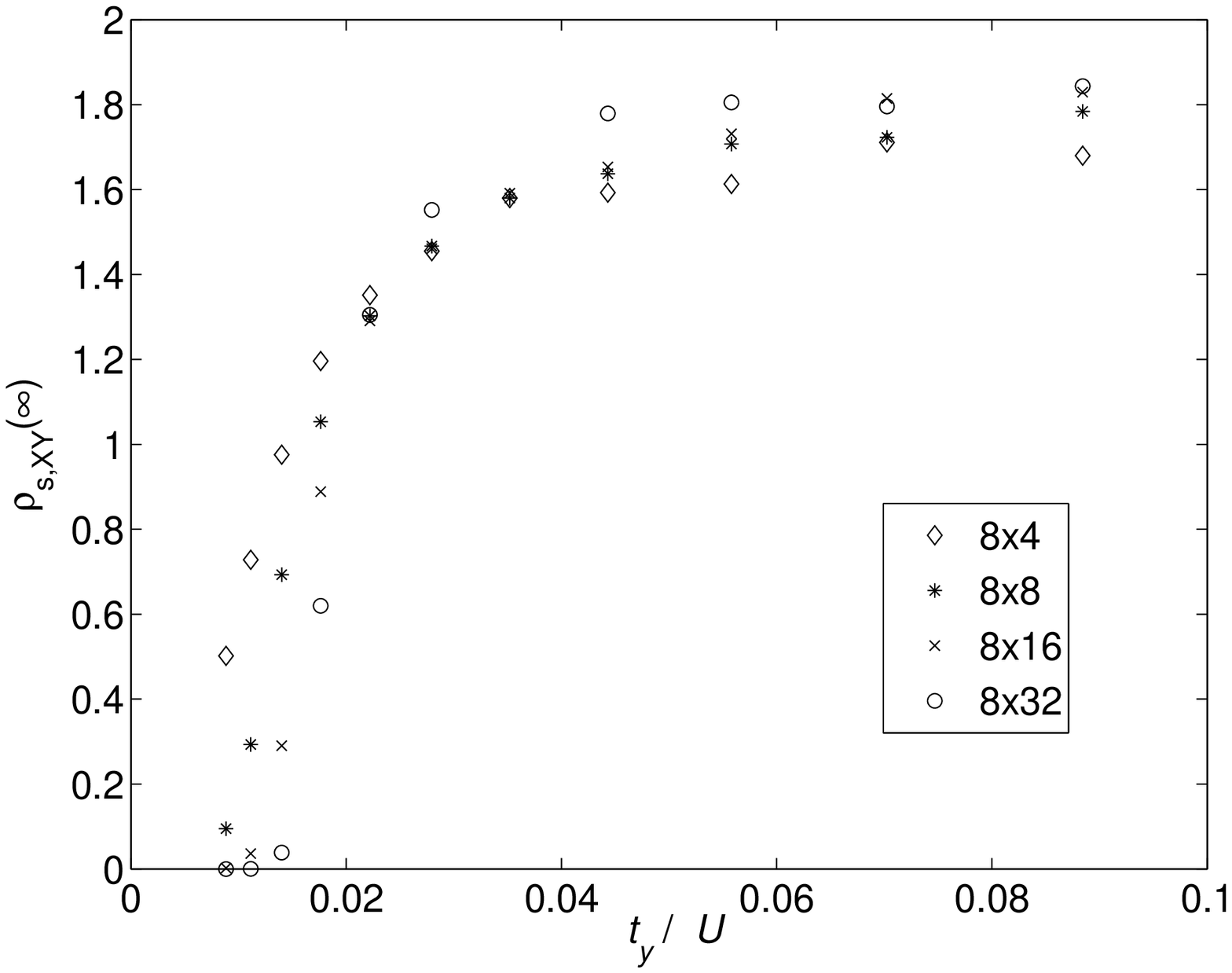}
\showfig{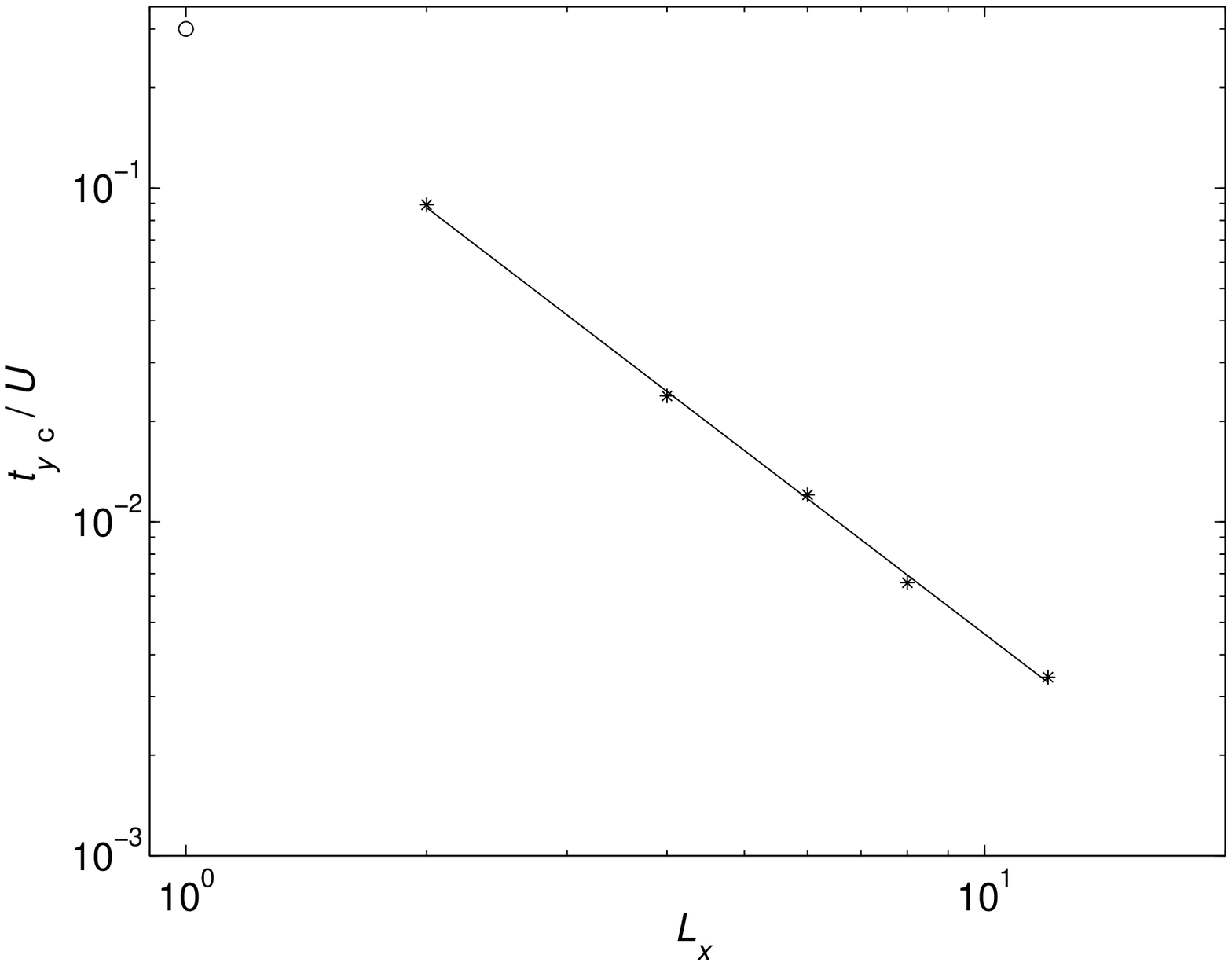}
\caption{Upper: An example of data collapse for the superfluid density 
scaled according to Weber-Minnhagen scaling, at side length $L_x=8$. 
Lower: Measured critical tunnelling as a function of side length $L_x$; 
the dependence is expected to be a power law. The linear fit yields 
$\alpha_x=0.17$, giving $K_x=2.9$.
The circle indicates the previously obtained critical point 
for the Mott transition in the 1D Hubbard model, $L_x=1$.
}
\label{fig:fittc}
\end{figure}

In order to check the above results, we apply a different scaling 
procedure, by bunching together results for the same $L_x$ but different 
$L_y$. 
If the superfluid density, corrected as in Eq.~(\ref{weberminnhagen}),
is computed for a range of $L_y$ values and a fixed $L_x$, the dependence 
on $L_x$ cancels out and the curves should 
%EL
%collapse 
coincide
at the critical point. 
The dependence of the critical point on $L_x$ can then be calculated. 
Figure \ref{fig:fittc} shows how this method works. 
The point $\tc/U$, at which the variance of the Weber-Minnhagen scaled 
superfluid density across different $L_y$ is a minimum, is recorded 
for each fixed value of $L_x$. (The result for $L_x=8$ is indicated with an 
arrow in the upper panel of Fig.\ \ref{fig:fittc}.) 
Then $\tc/U$ as a function of $L_x$ is 
fitted to a power-law dependence, as illustrated in the lower panel of
Fig.~\ref{fig:fittc}. 
The best linear fit to the log-log-curve 
is given by $t_c=0.31 L_x^{-1.83}$. 
In the figure, we have also inserted the previously obtained 
result $\ttc=0.30$ for 
the case $L_x=1$, which is just the 1D Hubbard model \cite{kuhner2000}.
%The spread in the data is 
%too large to allow for quantitative conclusions. 
%However, 
Fig.\ \ref{fig:fittc} is a qualitative support for the 
prediction of Refs.\ \thecitation that the critical coupling decreases 
as a power-law function of $L_x$, with a power slightly below 2.

\section{Phase transition and correlations}
\label{sec:correlations}
\begin{figure}
\showfig{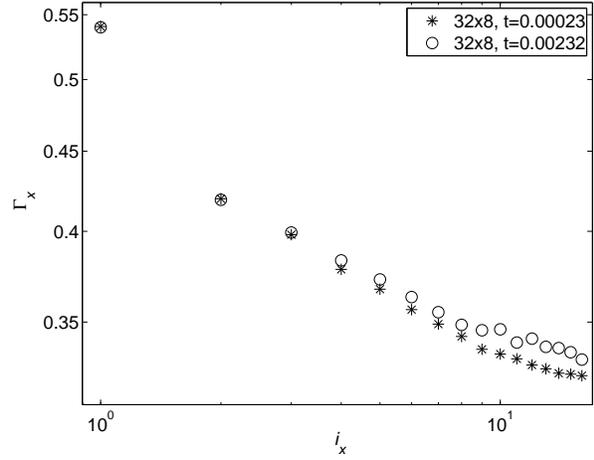}
\caption{Particle-particle correlation function in the $x$ direction. 
The correlation function $\Gamma_x$ is shown as a function of 
coordinate $i_x$ for filling $\rho=1$, $t_x/U=0.5$, 
$L_x=32$, $L_y=8$, $t_y/U=2.32\times 10^{-4}$ ($\ast$), and 
$t_y/U=2.32\times 10^{-3}$ ($\circ$).
}
\label{fig:correlation}
\end{figure}

One important prediction made in Refs.\ \thecitation is that the 
dependence on the transition point $\tc$ on tube length $L_x$ is 
linked to the behaviour of the particle-particle correlations 
in an isolated tube. In order to test this, we compute the correlation 
function $\Gamma_x(i_x)$, where $i_x$ is the number of lattice sites 
separating two points in the $x$ direction. 
An example of a computed correlation function 
is shown in Fig.~\ref{fig:correlation}. 
The correlation function is fitted to a power law according to 
\refeq{correlation}. 
It is seen in Fig.~\ref{fig:correlation} that the correlations in the 
strongly coupled $x$ direction depend on the tunnelling in the 
weakly coupled $y$ direction (just as the 
opposite relation holds). 
Since the predictions of Refs.~\thecitation build on the correlation 
properties of an isolated tube, we should use the results obtained for the 
smallest values of $t_y$, in the 1D MI phase, where the tubes are 
decoupled. We find the value $K_x=3.0$ for $t_x=0.5U$ and filling 
$\rho=1$. 
%where the error is the standard deviation over all different values of $t_y$.
This
is consistent with the value $K_x=3.4$ found from finite-size scaling 
in Sec.~\ref{sec:scaling}.

\section{Dependence on in-tube tunneling}
\label{sec:doubletransition}
\begin{figure}
\showfig{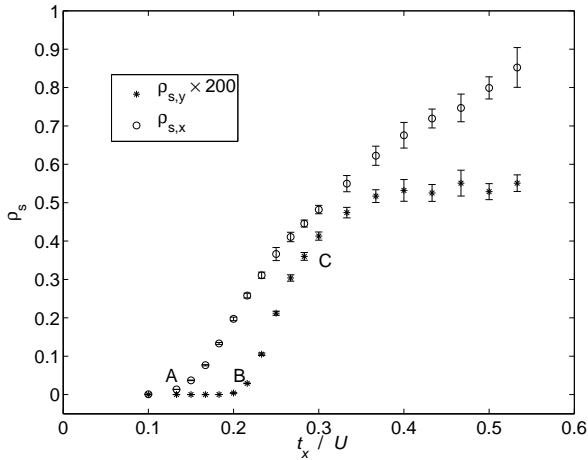}
\caption{Superfluid density in the $x$ and $y$ directions, respectively, 
for a Bose-Hubbard model of size 12$\times$12 lattice sites, fixed 
tunnelling $t_y/U=0.003$, and a filling of one atom per site. The scale 
for $\rho_{s,y}$ has been expanded by a factor 200. The points labelled 
A, B, and C are 
discussed in the text.}
\label{fig:horizontal}
\end{figure}

As a way to visualise the three predicted phases, we show 
as an example 
in 
Fig.\ \ref{fig:horizontal} the result of a calculation where the 
tunnelling in the strongly coupled $x$ direction, $t_x$, has been changed 
while $t_y$ is kept constant at $t_y=0.003U$. 
This corresponds to moving along a horizontal line in the lower part of 
the phase diagram in Fig.\ \ref{fig:phasediagram}. 
(Observe, however, that Fig.\ \ref{fig:phasediagram} corresponds to the 
case $L_x=4$, while here $L_x=12$, and therefore the position of the 
phase boundary is shifted.) 
The parameters are chosen such 
that, anticipating the results in Sec.\ \ref{sec:scaling}, 
the system should pass from the 2D MI phase, via the 1D MI phase, 
into the 2D SF phase.
The calculation is done for a finite lattice with 12$\times$12 sites, 
and the superfluid densities corresponding to motion in the $x$ and $y$ 
directions, respectively, are found. 
It is seen that the superfluid density corresponding to the $x$ direction 
begins to increase first, at the point labelled A, and the increase of 
$\rho_{s,y}$ commences at the later point B. This is
precisely what is expected 
for a system that crosses the two transition lines. 
However, one should note that the increase of $\rho_{s,y}$ seems to saturate 
at point C, and at the same point, the slope of the curve for $\rho_{s,x}$ 
is also seen to slightly decrease. 

In fact, when one 
tunes $t_x$ such that the 2D MI, 1D MI, and 2D SF phases 
are visited in turn, then both phase transitions belong to the BKT 
universality class. Furthermore, experience shows that it is the onset 
of the drop from a finite value, not the onset of a rise from a value 
close to zero, that should be identified with the BKT transition point. 
In Fig.\ \ref{fig:horizontal} it thus seems that it is the point labelled 
C, rather than the points labelled A and B, that indicates the true 
transition, and one 
can conclude that the transitions 
associated with the strong and weak couplings occur (within calculated 
error bars) at the {\it same} point. 
This is also what the theory for the {\it infinite} system 
dictates: there is no 1D MI phase 
if both $L_x$ and $L_y$ are taken to infinity. 
Here, however, we are concerned with {\it finite} systems, so we should 
not take the limit of infinite $L_x$. 
The figure shows clearly that in the finite system, there is a region in 
which the superflow along the $x$ direction is non-negligible but that 
along the $y$ direction is very small. This is the 1D MI phase, but 
it cannot be found by applying finite-size scaling for the $x$ direction.

\section{Phase transition at non-integer density}
\label{sec:densities}
\begin{figure}
\showfig{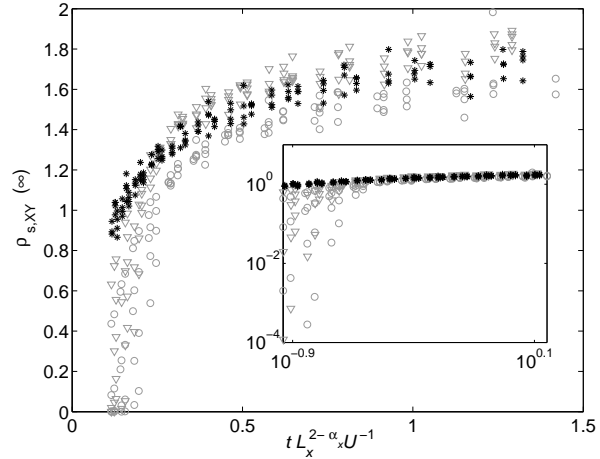}
\caption{Superfluid density in the XY model, $\rhoxy$, for a 
number of data sets. Data is taken at the commensurate 
density $\rho=1$ ($\circ$); commensurate tube filling with $\rho=5/4$ 
($\bigtriangledown$); and the incommensurate filling $\rho=19/16$ ($\ast$).
Here, $t_x/U=0.5$, the value of $L_x$ for the different curves 
varies between 2 and 32, and $L_y$ between 4 and 32. 
The quantities on the 
axes are the best fits to finite-size scaling for the coupling and 
superfluid density, respectively.
The inset shows the same data in a log-log plot.
}
\label{fig:globalcollapse}
\end{figure}

We return to Eq.~(\ref{transnphi}), obtained from Refs.~\thecitation, 
which predicts that the transition point $\tc/U$ has a power-law dependence 
on both the length $L_x$ and filling $\rho$ of the tubes. However, the 
1D MI phase only exists if there is an integer number of bosons in each 
tube, {\it i.e.}, if the filling is commensurate with respect to the number of 
{\em tubes}. The number of particles per {\em site}, on the other hand, 
does not have to be an integer. 
To check this, and thus establish that we are indeed seeing a Mott 
transition along one direction, we study the cases $\rho=5/4$ and 
$\rho=19/16$, respectively. In the first case, the tube length, $L_x$, 
is chosen as a multiple of 4 in order to ensure commensurability, and in 
the second case, we choose values of $L_x$ that are not divisible by 
16, in order to avoid commensurability.
The results are shown in Fig.~\ref{fig:globalcollapse}.

The finite-size scaling was performed as described in Sec.~\ref{sec:scaling}.
In the commensurate case, $\rho=5/4$, a phase 
transition is found. 
For the incommensurate case, $\rho=19/16$, the data may be collapsed with 
the best-fit result $K_x= 3.6$, and $\ttc=0.21$. 
However, it is seen 
in Fig.~\ref{fig:globalcollapse} that the superfluid density for this 
incommensurate density does not go 
steeply to zero when $t_y$ is 
decreased below the calculated critical point, even for the largest system size 
$L_y=32$. Instead, the data is clearly consistent with the curves meeting 
the $x$ axis at the origin, unlike the commensurate cases $\rho=1$ and 
$\rho=5/4$. We conclude that the data corroborates the conclusion that 
the filling per tube needs to be integer in order for the 1D MI phase to 
exist.

\section{Conclusions}
\label{sec:conclusion}
%Anders version
In this paper, we have studied bosons trapped in two-dimensional optical lattices by Monte-Carlo calculations, with the objective of characterising phase transitions and their dependence on dimensionality and lack of isotropy. Apart from the expected phases, where one have superfluidity or Mott insulation along both directions, we show that there also can exist situations where atoms may tunnel along one direction, while not along the other. This means having superfluidity, and accordingly strong correlations, in only one of the two available dimensions. We call this phase a one-dimensional Mott insulator.

We study the transition to this phase from a two-dimensional superfluid and explore the conditions for this phase transition to occur. We find that the transition point depends on a specific combination of the weaker tunnelling matrix element $t_y$,  the on-site interaction strength $U$, the number of sites in the strongly coupled direction $L_x$, the filling $\rho$,
and in addition the Luttinger parameter $K_x$, which depends on the stronger tunnelling matrix element $t_x$. The transition point and the Luttinger parameter are both calculated.

We also verify that the location of the phase transition is connected to the decay of particle-particle correlations in a manner consistent with predictions based on Tomonaga-Luttinger liquid theory \thecitation, and that the transition occurs when the number of particles is commensurate with the side length of the system in the direction of
weak tunnelling, but not necessarily with the number of sites.

\begin{acknowledgement}

This work was supported by the G\"oran Gustafsson foundation, the
Swedish Research Council, the Knut and Alice Wallenberg Foundation,  
the Carl Trygger foundation, SIDA/SAREC, and the Kempe foundation. 
This research was conducted using the resources of High Performance 
Computing Center North (HPC2N).
M.R., R.S., M.Z., A.K., and E.L.\ are grateful to 
Mats Nyl\'en 
and 
Peter Olsson for helpful discussions. 
\end{acknowledgement}

%\bibliographystyle{../../../../bibfiles/prsty}
%\bibliography{../../../../bibfiles/shortnames,../../../../bibfiles/articles,../../../../bibfiles/books}

\end{document}